# Study of Migration of Giant Planets and Formation of Populations of Distant Trans-Neptunian Objects in the Nice Model

V. V. Emel'yanenko

*Institute of Astronomy, Russian Academy of Sciences, Moscow, Russia*
*\*e-mail: vvemel@inasan.ru*

Numerical modeling of the interaction of giant planets and the planetesimal disk was carried out for the Nice model, in which the initial orbits of the planets are in resonant configurations. In addition to the standard Nice model, planetesimals in the planetary region were considered and the self-gravity of the planetesimal disk was taken into account. The dynamical evolution of planetary systems has been studied for time intervals on the order of the lifetime of the Solar System. We have found cases in which the planetary systems survive for billions of years, the final orbits of the planets are close to the present orbits, and distant trans-Neptunian objects exist.

## INTRODUCTION

The discovery of numerous exoplanet systems and the detection of a population of trans-Neptunian objects with complex structures has had a profound impact on modern views of the formation of the Solar System. One of the most important advances was the understanding that the giant planets migrated significantly from their formation positions as a result of interactions with the planetesimal disk left after the dissipation of gas.

This is most expressed in the Nice model (Tsiganis et al., 2005; Morbidelli et al., 2007; Batygin and Brown, 2010; Levison et al., 2011; Nesvorný and Morbidelli, 2012). The recent discovery of a family of distant trans-Neptunian objects (see, for example, the review (Gladman, Volk, 2021)), moving in orbits with perihelion distances $q>40$ au and semi-major axes $a>150$ au, provided new and rather unexpected information about the structure of the outer Solar System, which became the basis for putting forward the hypothesis of the existence of a distant ninth planet (Trujillo, Sheppard, 2014; Batygin, Brown, 2016). Naturally, the question arises to what extent the Nice model is consistent with the existence of distant trans-Neptunian objects.

The work (Emel'yanenko, 2022) showed that distant trans-Neptunian objects are a natural result of the long-term evolution of a system that includes migrating giant planets and a self-gravitating planetesimal disk. In particular, this work was able to explain the origin of Sedna-type objects. The results obtained were related to a model in which the dynamical evolution of giant planets and the disk of planetesimals, initially located outside the planetary system, was considered. The initial conditions for the planets and the disk of planetesimals were similar to those considered in the papers (Kaib and Sheppard, 2016; Nesvorný et al., 2016), aimed at explaining the existence of orbits with large perihelion distances in the so-called "scattered disk" of trans-Neptunian objects. In fact, these works are related to the study of the final stage in the Nice model. It is natural to assume that a certain proportion of planetesimals remained after the formation of planets within the planetary region too. In particular, the paper (Silsbee and Tremaine, 2018) studied the dynamics of planetary embryos located initially between the planets, and it was shown that these objects can move to orbits located far beyond the planetary region. In the work (Silsbee and Tremaine, 2018), the planets were located near the present orbits, and their migration was not taken into account.

In the present work, we try to take into account the possible influence of planetesimals in the planetary region on the dynamical process of the formation of distant trans-Neptunian objects. In new numerical experiments, we examine the gravitational interaction of planets with planetesimals located initially both outside the planetary region and between planets. In such models, the total mass of the planetesimal disk may be greater than in the work (Emel'yanenko, 2022), where planetesimals located outside the planetary region were considered. This is an important circumstance, since there is a tendency for the number of distant trans-Neptunian objects produced to increase with increasing initial mass of the planetesimal disk. The initial planetary configurations, previously studied within the framework of the Nice model, are considered (Batygin et al., 2011; Nesvorný and Morbidelli, 2012). Planetary migration is a significant factor in these models.

The papers (Batygin et al., 2011; Nesvorný and Morbidelli, 2012) did not take into account the gravitational interaction of planetesimals. In addition, these papers considered relatively short time intervals (500 and 100 Myr, respectively). Naturally, distant trans-Neptunian objects were not found in these papers, which begin to appear in the self-gravitating planetesimal disk after several hundred million years (Emel'yanenko, 2022). In the present work, we analyze these models by considering the full *N*-body problem for the age of the Solar System.

## METHODS

We consider the gravitational interaction of bodies in systems consisting of four giant planets with their present masses and a large number of planetesimals with significantly lower masses. In this work, we studied two initial configurations of planets from the work (Nesvorný and Morbidelli, 2012): the most compact, in which Jupiter, Saturn, Uranus, and Neptune were in resonances 3 : 2, 3 : 2, and 4 : 3, respectively, and the most extended, in which these planets were in the corresponding resonances 3 : 2, 2 : 1, and 2 : 1. In addition, the initial configuration of planets located in resonances 2 : 1, 4 : 3, and 4 : 3 was considered, which best reproduces the dynamical structure of the Kuiper belt according to the study (Batygin et al., 2011). The initial parameters of the planetesimal disk (mass, boundaries) were chosen in such a way that after four billion years the planets would be located in orbits close to modern ones, based on previous experience in modeling planet migration (for example, Nesvorný and Morbidelli, 2012; Emel'yanenko, 2022). In particular, during the migration process, Jupiter and Saturn must overcome the 2:1 resonance and in the end be located near the present mean motion commensurability of 5 : 2. Naturally, due to the stochastic nature of the process of interaction of planets with planetesimals, for each initial configuration it was necessary to carry out a series of integrations with values of the disk parameters from a certain range. More detailed characteristics of the initial disk parameters are presented below for each case.

At the first stage, to create resonant configurations, we used a widely used technique (for example, Batygin and Brown, 2010; Nesvorný and Morbidelli, 2012;Clement et al., 2021a, 2021b). Initially, the orbits of the planets were located somewhat further than the resonant positions. The planets then migrated inward through the introduction of additional nongravitational forces are captured in resonances (Papaloizou and Larwood, 2000; Emel'yanenko, 2011). The integration time at this stage ranged from 180 thousand years to 630 thousand years for different variants. In all variants, the orbits obtained at the end of this stage had eccentricities $e < 0.07$ and inclinations $i < 0.15°$. The planetesimal disk was represented by 1000 objects, of which a certain part (from 170 to 500 objects) had the same nonzero mass. The initial values of eccentricities and inclinations of planetesimal orbits were distributed uniformly in the intervals (0, 0.01) and (0°, 0.5°), respectively. Semimajor axes of planetesimals $a$ were distributed according to a power law $a^{-s}$, where $s$ took values from 0 to 1.5.

A numerical solution of the equations of motion in the *N*-body problem was carried out for 4 Gyr, using the symplectic integrator (Emel'yanenko, 2007). Objects were removed from integration if $a > 2500$ au or $e > 1$ away from perturbing bodies. We also removed objects passing at a distance of less than 0.1 au from the Sun. We stopped the integration if the planetary system was destroyed. In this case, a new integration was launched with the same system parameters, but a new random distribution of the initial orbits of planetesimals. As a rule, the maximum number of integration runs for a system with the same parameters was ten. In the most interesting cases, we carried out integration for a larger number of simulations.

## RESONANCE 3 : 2, 3 : 2, 4 : 3

In this variant, at the initial moment, all planetesimals were located beyond the orbit of Neptune, since it is difficult to assume that objects could remain in stable orbits between the giant planets in such a compact system. The number of massive planetesimals was 170, as in (Emel'yanenko, 2022). The characteristics of the systems under consideration are given in Table 1 ($a_N$ is the initial value of the semimajor axis of Neptune's orbit, the semimajor axes of planetesimals in the initial disk with mass $M_d$ were distributed between $a_{in}$ and $a_{out}$).

**Table 1.** Characteristics of the studied systems with initial giant planets located in resonances 3 : 2, 3 : 2, 4 : 3

| $M_d$, Earth mass | $a_N$, au | s | $a_{in}$, au | $a_{out}$, au |
|---|---|---|---|---|
| 20 | 11.43 | 0.5, 1.0, 1.5 | 11.93 | 37.0 |
| 40 | 12.08 | 0.5, 1.0, 1.5 | 12.58 | 32.0 |
| 50 | 12.41 | 0.5, 1.0, 1.5 | 12.91 | 30.0 |
| 60 | 12.54 | 0.5, 1.0, 1.5 | 13.04 | 29.0 |
| 70 | 12.69 | 0.5, 1.0, 1.5 | 13.19 | 28.0 |

Calculations showed that the disk mass $M_d = 20 M_E$, where $M_E$ is the mass of the Earth, in all the cases considered, is clearly insufficient to transfer the planets from the initial resonant configuration to the present orbits.

For more massive disks with $M_d \geq 40 M_E$ the degree of migration is sufficient to transfer planets to the present orbits. However, in most of the cases considered, the planetary systems turned out to be unstable; the planets remained in almost circular orbits for no more than 360 Myr. Only two cases have been recorded in which the planetary system survived for 4 Gyr. The first case is when $M_d = 40 M_E$ and $s = 0.5$, but here the final ratio of the periods of Saturn and Jupiter was only 1.84. In the second case when $M_d = 60 M_E$ and $s = 1.0$, the indicated ratio is quite acceptable (2.48), but the final value of Jupiter's eccentricity was only 0.001, which is in clear contradiction with the present value. In this variant, only one distant trans-Neptunian object with the perihelion distance $q = 63$ au and $a = 162$ au remained.

In total, we can say that the initial resonant configuration of the planets 3 : 2, 3 : 2, 4 : 3 does not favor the creation of a planetary system with distant trans-Neptunian objects. It is impossible, of course, to say categorically that for the disk parameters that led to instability in our calculations, the desired systems cannot be realized after 4 Gyr of evolution. But the probability of such an event for a given initial configuration is small.

### RESONANCE 3 : 2, 2 : 1, 2 : 1

In this variant, at the initial moment, the planetesimals were located between the orbit of Jupiter and $a_{out}$ (Table 2), but within two Hill radii relative to the semimajor axes of the planets, planetesimals were absent (see Silsbee and Tremaine, 2018). The number of massive planetesimals was 500. The characteristics of the systems under consideration are given in Table 2. In the case of the initial mass of the disk $M_d = 40 M_E$ no satisfactory cases were found in our simulations. Planetary systems either became unstable in less than 150 Myr, or the ratio of the periods of Saturn and Jupiter was significantly less than the present value.

More satisfactory results were obtained with $M_d = 60 M_E$. In many cases, planetary orbits are obtained that are close to the present orbits of the giant planets, but in a relatively short period of time (less than 200 Myr). As the planets continue to migrate, such systems become unstable. But for $s = 1.0$, cases were found in which the planets survive near the present orbits for more than 2 Gyr, and distant trans-Neptunian objects exist. Because of the importance of this case, it is discussed in more detail below in a separate section. For a more massive disk with $M_d = 80 M_E$ there are many cases with stable planetary systems and distant trans-Neptunian objects. But in these cases there is a significant difficulty associated with the relative position of the orbits of Jupiter and Saturn. The migration of these planets is too fast, so that in all studied cases the mean motion ratio for Jupiter and Saturn begins to exceed 2.5 in less than 200 Myr. Subsequently, this ratio continues to increase, and in all systems that survive for several billion years, the semimajor axis of Saturn's orbit exceeds 10 au.

**Table 2.** Characteristics of the studied systems with initial giant planets located in resonances 3 : 2, 2 : 1, 2 : 1

| $M_d$, Earth mass | $a_N$, au. | s | $a_{out}$, au. |
|---|---|---|---|
| 40 | 18.91 | 0, 0.5, 1.0 | 29.9 |
| 60 | 19.48 | 0, 0.5, 1.0 | 28.6 |
| 80 | 19.89 | 0, 0.5, 1.0 | 27.4 |

RESONANCE 2 : 1, 4 : 3, 4 : 3

As in the previous case, in this version, at the initial moment, the planetesimals were located between the orbit of Jupiter and $a_{out}$ (Table 3), but within two Hill radii relative to the semimajor axes of the planets there were no planetesimals. The number of massive planetesimals was 500. The characteristics of the systems under consideration are given in Table 3.

**Table 3.** Characteristics of the studied systems with initial giant planets located in resonances 2 : 1, 4 : 3, 4 : 3

| $M_d$, Earth mass | $a_N$, au | s | $a_{out}$, au |
|---|---|---|---|
| 40 | 13.29 | 0, 0.5, 1.0 | 29.9 |
| 60 | 13.63 | 0, 0.5, 1.0 | 28.6 |
| 80 | 13.89 | 0.5 | 27.4 |

For the initial value $M_d = 40 M_E$ in our simulations, cases were found in which the planetary system survives for more than 2 Gyr and distant trans-Neptunian objects exist. In one version (with $s = 0.5$) the planetary system and three trans-Neptunian objects, one of which is distant, remain after 4 Gyr. But at the same time, the ratio of the periods of Saturn and Jupiter exceeds 2.5. For $s = 0$, a case was found in which the system survives for more than 2 Gyr with the ratio of the periods of Saturn and Jupiter close to the modern value of 2.49.

For more massive disks with $M_d = 60 M_E$ and $80 M_E$, the ratio of the periods of Saturn and Jupiter very quickly (in all cases less than 50 Myr) begins to exceed the modern value. Then either the planetary system is destroyed, or it differs significantly from the present system of giant planets.

DISCUSSION

Naturally, in conditions of long-term stochastic motion of planets and massive planetesimals, it is very difficult to select initial parameters under which, at a certain moment of time, a system will be realized in which the orbits of the planets will be close to the present orbits. In addition, the time interval during which the migration of planets occurred is uncertain. In our study, numerical experiments were carried out for an interval of 4 Gyr. But even in the Nice model this magnitude is poorly defined. An early version of this model suggested that planetary instability began to develop after several hundred million years of being in a resonant configuration, linking the moment of instability to the period of late heavy bombardment of the Moon (Morbidelli et al., 2007; Levison et al., 2011). Currently, preference is given to models in which the instability of the outer planet system began no later than several tens of millions of years after the gas dissipation in the protoplanetary disk (see, for example, Nesvorný, 2018).

Therefore, in Table 4 we collected information about those cases in which the ratio of the periods of Saturn and Jupiter reached the present value of 2.49 at some moment of time exceeding 2 Gyr. These systems are the most preferred for assessing the possible evolution of the orbits of planets, since the location of the planets, close to the present one, arises in them as a result of a very long-term gravitational interaction of the planets and the planetesimal disk. In this table $M_{od}$ is the total mass of planetesimals initially located beyond the orbit of Neptune, $t_f$ is the epoch at which the ratio of the periods of Saturn and Jupiter reached 2.49, and $M_{d,f}$ is mass of the planetesimal disk at the moment $t_f$.

**Table 4.** Characteristics of systems in which the giant planets move in orbits close to the present ones at some moment of time exceeding 2 Gyr

| Resonance | $M_d$, Earth mass | $M_{od}$, Earth mass | $s$ | $t_f$, Gyr | $M_{d,f}$, Earth mass |
|---|---|---|---|---|---|
| 3:2, 2:1, 2:1 | 60 | 18 | 1.0 | 2.45 | 0.72 |
| 2:1, 4:3, 4:3 | 40 | 33 | 0 | 2.10 | 0.48 |

Figure 1 shows the change in the semimajor axes of the planetary orbits for the 2 : 1, 4 : 3, 4 : 3 variant, and Fig. 2 shows the distribution of semimajor axes and perihelion distances for planetesimals remaining at the time $t_f$. The semimajor axes of Neptune and Uranus do not perfectly correspond to the present values, but this is caused only by a choice of the outer boundary of the initial planetesimal disk. As can be seen, in this version there are distant trans-Neptunian objects.

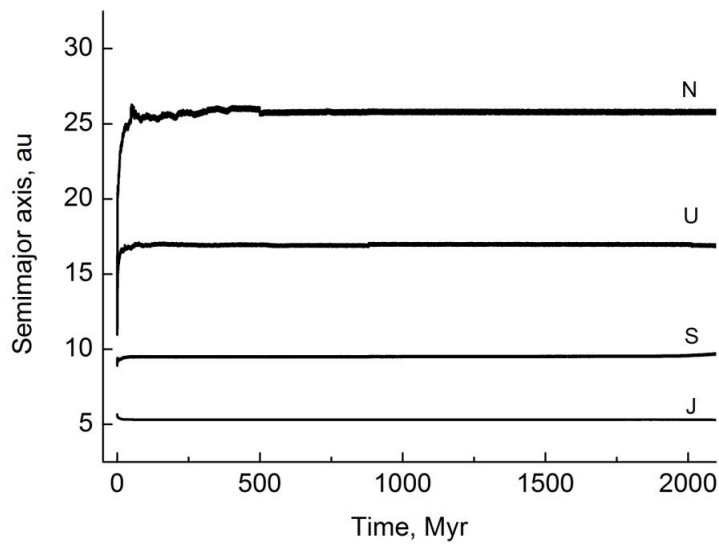

**Fig. 1.** Changes in the semimajor axes of the orbits of planets that started in resonance 2 : 1, 4 : 3, 4 : 3 (J—Jupiter, S—Saturn, U—Uranus, N—Neptune).

Figure 3 shows the change in the semimajor axes of the planetary orbits in the 3 : 2, 2 : 1, 2 : 1 variant, and Fig. 4 shows the distribution of semimajor axes and perihelion distances for planetesimals remaining at the moment $t_f$. It can be seen that in this version the orbits of the planets are more consistent with the configuration of the present planetary system than in the previous case. In the variant 3 : 2, 2 : 1, 2 : 1 there are also distant trans-Neptunian objects at the moment $t_f$.

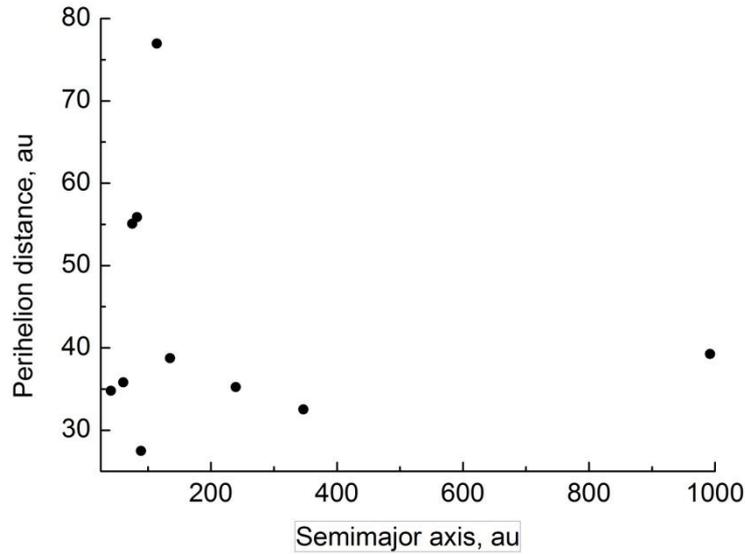

**Fig. 2.** Semimajor axes and perihelion distances of planetesimals at the moment $t_f$ in the variant 2 : 1, 4 : 3, 4 : 3.

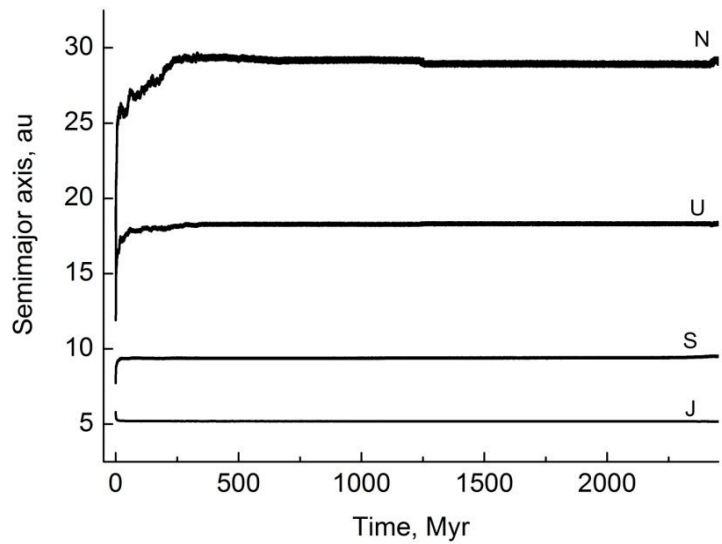

**Fig. 3.** Changes in the semimajor axes of the orbits of planets that started in resonance 3 : 2, 2 : 1, 2 : 1 (J—Jupiter, S—Saturn, U—Uranus, N—Neptune).

Figure 5 illustrates how objects transfer to distant orbits. Here is an example of changing perihelion distance, semimajor axis and longitude of perihelion $\pi$ for a massive object reaching an orbit with $q = 48$ au, $a = 326$ au at the moment $t_f$ (lines with the number *1*). During most of its evolution, the object moves in an orbit with a perihelion distance of about 40 au. But over the time interval from 2.0 Gyr to 2.2 Gyr, the rate of change in the longitude of the perihelion of this object becomes close to the rate of change of the longitude of perihelion for another massive object (changes in the orbital elements for the second object are shown in Fig. 5 by lines with the number *2*). Thus, these objects move in a secular resonance. In this situation, the perihelion distances of objects change significantly. The perihelion distance of the first object increases, and the perihelion of the second object enters the planetary region. Subsequently, the second object is thrown into the region $a > 2500$ au due to perturbations from the giant planets, and the first object remains in the distant trans-Neptunian region.

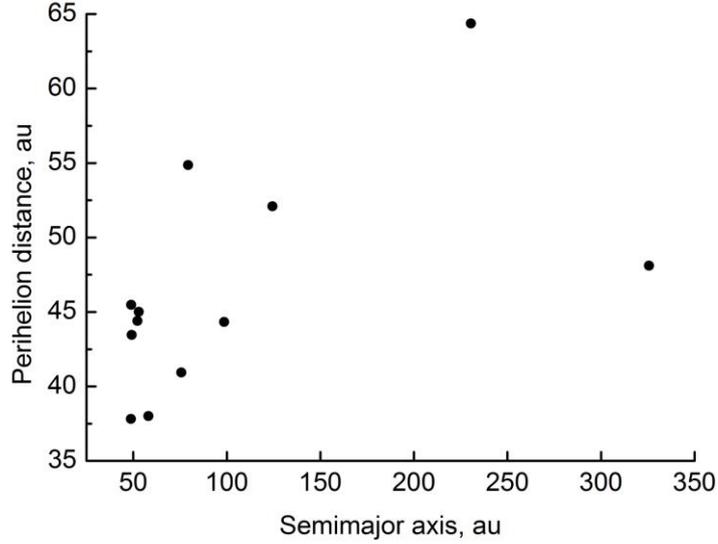

**Fig. 4.** Semimajor axes and perihelion distances of planetesimals at the moment $t_f$ in the variant 3 : 2, 2 : 1, 2 : 1.

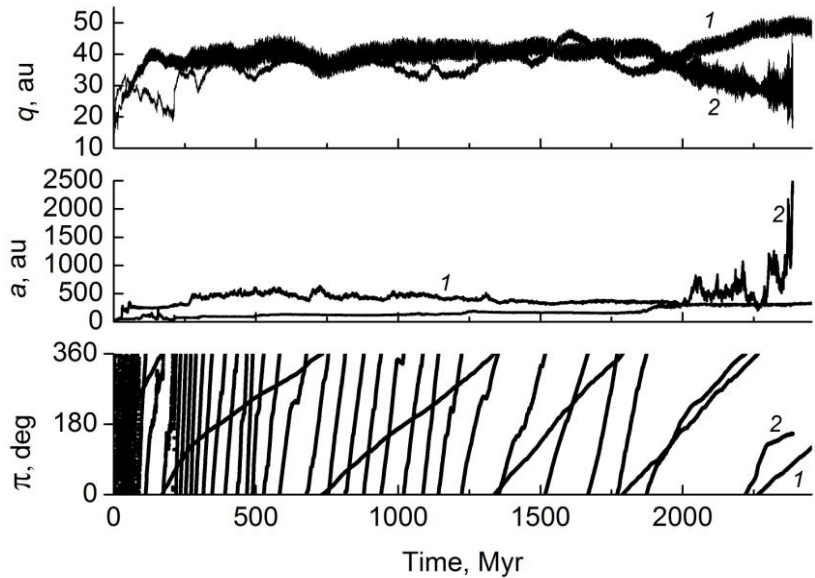

**Fig. 5.** Changes in perihelion distance $q$, semimajor axis $a$, and longitude of perihelion $\pi$ for two massive planetesimals.

CONCLUSIONS

Simulations of the interaction of the giant planets and the planetesimal disk were carried out for the Nice model, in which the initial orbits of the planets are in resonant configurations. In addition to the standard Nice model, planetesimals in the planetary region were considered and the self-gravity of the planetesimal disk was taken into account. The dynamical evolution of planetary systems has been studied over time intervals on the order of the lifetime of the Solar System.

In most numerical simulations within the Nice model, either planetary systems are destroyed or planets are transferred to orbits that differ significantly from the present orbits. But, if the system survives for billions of years, then, as a rule, it has distant trans-Neptunian objects. We have found cases with $M_d = 60 M_E$ for resonance 3 : 2, 2 : 1, 2 : 1 and $M_d = 40 M_E$ for resonance 2 : 1, 4 : 3, 4 : 3, in which the planetary systems survive for billions of years, the final orbits of the planets are close to the present orbits, and distant trans-Neptunian objects exist.

## ACKNOWLEDGMENTS
The calculations were carried out using the MVS-10P supercomputer of the Joint Supercomputer Center of the Russian Academy of Sciences.